\documentclass[aps,prd,twocolumn,nofootinbib,citeautoscript,showkeys]{revtex4-1}        
\usepackage{amsmath,amssymb,mathrsfs,bm,setspace}
\usepackage{graphicx}  
\usepackage[tight]{subfigure}     
\usepackage{color} 
\usepackage{braket}
\usepackage{array}
\usepackage[all]{xy}
\usepackage{float}

\usepackage[colorlinks=true]{hyperref}  
\hypersetup{
    bookmarks=true,         
    unicode=false,          
    pdftoolbar=true,        
    pdfmenubar=true,        
    pdffitwindow=false,     
    pdfstartview={FitH},    
    pdftitle={My title},    
    pdfauthor={Author},     
    pdfsubject={Subject},   
    pdfcreator={Creator},   
    pdfproducer={Producer}, 
    pdfkeywords={keyword1} {key2} {key3}, 
    pdfnewwindow=true,      
    colorlinks=true,       
    linkcolor=black, 
    citecolor=blue,        
    filecolor=magenta,      
    urlcolor=cyan           
} 
\definecolor{darkblue}{rgb}{0.2, 0, 0.8}
\definecolor{darkgreen}{rgb}{0.2, 0.71, 0}

\numberwithin{equation}{section}

\newcommand{\be}{\begin{equation*}}
\newcommand{\ee}{\end{equation*}}
\newcommand{\ben}{\begin{equation}}
\newcommand{\een}{\end{equation}}
\newcommand{\beqa}{\begin{eqnarray*}}
\newcommand{\eeqa}{\end{eqnarray*}}
\newcommand{\beqan}{\begin{eqnarray}}
\newcommand{\eeqan}{\end{eqnarray}}
\newcommand{\nn}{\nonumber}

\newcommand{\R}{\mathcal{R}}
\newcommand{\Z}{\mathcal{Z}}

\renewcommand{\href}[2]{#2}

\def\Sp{\mathrm{Sp}}

\def\Aut{\mathrm{Aut}}

\def\R{\mathbb{R}}
\def\Z{\mathbb{Z}}
\def\eqdef{\stackrel{\mathrm{def}}{=}}
\def\U{\mathrm{U}}

\def\SO{\mathrm{SO}}

\def\Spin{\mathrm{Spin}}
\def\Pin{\mathrm{Pin}}
\def\End{\mathrm{End}}
\def\Cl{\mathrm{Cl}}
\def\L{\mathrm{L}}
\def\cL{\mathcal{L}}
\def\Ad{\mathrm{Ad}}
\def\O{\mathrm{O}}
\def\ev{{\mathrm{ev}}}
\def\w{\mathrm{w}}
\def\fG{\mathfrak{G}}

\newcommand{\twopartdef}[4]
{
	\left\{
		\begin{array}{ll}
			#1 & \mbox{if } #2 \\
			#3 & \mbox{if } #4
		\end{array}
	\right .
}

\begin{document}

\title{On the spin geometry of supergravity and string theory}  
\author{C. I. Lazaroiu and C. S. Shahbazi}\vspace{0.1cm}
\affiliation{Center for Geometry and Physics, Institute for Basic
  Science, Pohang, Republic of Korea 37673 } 
\email{calin@ibs.re.kr} 
\affiliation{Institut de Physique Th\'eorique, CEA-Saclay, France.}
\email{carlos.shabazi-alonso@cea.fr}

\date{\today}

\begin{abstract}     
We summarize the main results of our recent investigation of bundles of 
real Clifford modules and briefly touch on some applications to string theory and
supergravity.
\end{abstract}   

\maketitle \singlespacing
  

\section{I\lowercase{ntroduction}} 
\label{sec:intro}  


Spinor fields are a crucial ingredient of supersymmetric theories such
as superstring theory and supergravity. As such, a complete
understanding of spin geometry in all dimensions and signatures for
which such theories can be defined is of paramount importance for
determining the weakest assumptions under which one can make global
sense of various models. Virtually all supergravity Lagrangians
currently in use are only locally defined, and the task of promoting
them to complete formulations of the corresponding theories is
affected by ambiguities in how the local formulas should be
interpreted globally. Some of the most subtle issues of this type
concern the global nature of spinorial fields, whose precise
determination requires, in particular, that one identifies the weakest
topological conditions which guarantee that a particular global
interpretation is in fact allowed. The problem is compounded in string
theory, which induces quantum corrections whose precise nature is
sensitive to the global spin geometry of the target space-time $(M,g)$, D-brane
world-volumes etc in various dimensions.

When seriously approaching the problem of giving global formulations
of supergravity and string theories, one soon finds out that the
questions: \emph{What is the allowed global nature of spinor fields in
  string and supergravity theories and what are weakest conditions
  under which such theories can be defined ?} remains to be fully
answered.

Locally, spinor fields in supergravity theories are usually considered
as functions valued in a vector space $S_0$ acted on by
locally-defined gamma matrices $\gamma^\mu$ associated to a vielbien
(=local frame of the tangent bundle of the space-time $M$) $e^\mu$.
This amounts to taking $S_0$ to be a representation of the Clifford
algebra $\Cl(T_x M, g_x)$ (or sometimes of its even part
$\Cl^\ev(T_xM,g_x)$) of the tangent space $T_xM$ at every point $x$ of
$M$. Since the formulation found in the literature is only local, the
vielbein and in fact the entire Lagrangian density is defined only on
a contractible open subset $U$ of $M$ and these Clifford algebras are
fibers of the restricted Clifford bundle $\Cl(TU,g|_U)$ (or of its
even sub-bundle $\Cl^\ev(TU,g|_U)$), both of which are topologically
trivial bundles of unital and associative $\R$-algebras. This local
set up does not determine uniquely the global nature of the spinor
fields appearing in a putative global extension of the local
Lagrangian densities found in the literature. A further complication
concerns the treatment of reality conditions. Although the local
structure of supergravity theories often requires that various spinor
fields are real, the physics literature dealing with these situations
usually starts by considering spinors as locally-defined functions
valued in a complex Clifford representation which is then
\emph{realified} by using an appropriate antilinear conjugation
operator, a generally non-canonical process that obscures the global
properties of the spinors. In particular, it sometimes happens that
the charge conjugation operator which one would wish to use in order
to impose a reality condition on a complex spinor field {\em cannot}
be defined globally as an appropriate section of the bundle of
endomorphisms of a spinor bundle, which means that the usual approach
found in the physics literature is often not appropriate for a global
and general description of real spinor fields.

Globally understanding spinor fields requires that one fully
characterizes the vector bundle over space-time of which they are
sections. A simple way to globalize Clifford representations is to
assume that the space-time admits a $\Spin$ structure $Q$. In that
case, spinor fields can be defined as global sections of a vector
bundle $S$ over $M$ which is associated to $Q$ through some given
representation of the spin group.  Such bundles $S$ carry a
globally-defined Clifford multiplication which makes them into bundles
of modules over the fibers of the Clifford bundle $\Cl(TM,g)$ of
space-time.

However, a spin structure has no immediate physical meaning and it is
not clear in general that one is physically required to assume that
$S$ is a vector bundle associated to such a structure. What is
physically important (and directly measurable) is not the spin
structure $Q$ itself but physical observables constructed using
sections of the vector bundle $S$. For example, one generally {\em
  cannot} deduce that $(M,g)$ must admit any spin structure solely
from the fact that it admits a vector bundle with a globally-defined
Clifford multiplication ! Careful reasoning along these lines shows
that some supergravity theories (and especially those coupled to
scalar and vector matter) can sometimes be defined globally {\em
  without} requiring that the space-time admits a spin
structure. Therefore, one is naturally lead to ask what are the {\em
  weakest} conditions under which a given theory with a locally known
Lagrangian density can be globally defined, and in how many
globally-inequivalent ways.

In fact, Clifford multiplication is the minimal ingredient needed to
construct a Dirac operator and hence to formulate the fermion kinetic
action. Hence, one should take Clifford multiplication and not spin
structures (or any of their known extensions) as the {\em a priori}
fundamental ingredient needed to describe the physics of spinor
fields. This reasoning suggests that one should define spinors not as
sections of a vector bundle associated to some $\Spin$ structure, but
as sections of an abstract vector bundle equipped with Clifford
multiplication. At this point, one has a choice between {\em inner}
and {\em outer} Clifford multiplications\footnote{Outer Clifford
  multiplication arises, for example, in the theory of Pin structures,
  in which situation it sometimes allows one to define a ``modified''
  Dirac operator \cite{ModifiedDirac}.}. The former is a morphism
$TM\otimes S\rightarrow S$ obeying the Clifford property while the
latter is a morphism $TM\otimes S\rightarrow S'$, where $S'$ is a
vector bundle which need not be isomorphic with $S$. In this note, we
shall consider exclusively bundles endowed with inner Clifford
multiplication, hence $S$ will be a bundle of modules over the full
Clifford bundle $\Cl(TM,g)$ of space-time. The fiber $S_p$ of $S$ at
every point $p$ of $M$ carries a representation of the algebra
$\Cl(T_pM,g_p)$. This gives a unital morphism of bundles of algebras
$\gamma:\Cl(TM,g)\rightarrow \End(S)$, which we shall call the {\em
  structure morphism} of $S$. For technical reasons, we also require
that $\gamma$ be {\em weakly faithful}, which means that the
restriction of $\gamma$ to the vector bundle $TM\subset \Cl(TM,g)$ is
injective. For brevity of language, we define a \emph{pinor
  bundle}\footnote{The word ``pinor'' refers to the fact that we
  consider bundles of modules over the fibers of $\Cl(TM,g)$ rather
  than over the fibers of $\Cl^\ev(TM,g)$.} to be a weakly faithful
bundle $S$ of Clifford modules. This definition leads to a few natural
questions:

\begin{itemize}
\item \emph{Is every pinor bundle associated to a spin structure? If
  not, to what principal bundle is it associated ?}
\item \emph{What is the topological obstruction to existence of a
  pinor bundle on a pseudo-Riemannian manifold $(M,g)$ of arbitrary signature
  $(p,q)$ ? }
\end{itemize}

These questions were addressed in \cite{Lipschitz}. Regarding
the first question, the results of loc. cit. show that, in general,
real pinor bundles are associated not to spin structures but to more
general spinorial structures, which we call \emph{real Lipschitz
  structures}, following previous work by T. Friedrich and A. Trautman
\cite{FriedrichTrautman} concerning the complex case. The second question 
was completely solved in \cite{Lipschitz} for so-called {\em elementary pinor bundles}, 
defined as those pinor bundles whose fiberwise Clifford representation is irreducible. 

\section{R\lowercase{eal} L\lowercase{ipschitz structures and their relation to real pinor bundles}}

Let $(V,h)$ be a quadratic vector space which is isomorphic with each
fiber of the tangent bundle $(TM,g)$. A representation
$\eta:\Cl(V,h)\rightarrow \End_\R(S_0)$ of the Clifford algebra
$\Cl(V,h)$ in a finite-dimensional real vector space $S_0$ is called
{\em weakly faithful} if the restriction of $\eta$ to the subspace $V$
of $\Cl(V,h)$ is injective. The {\em real Lipschitz group $\L(\eta)$}
of $\eta$ is the group consisting of all invertible operators $a$
acting in $S_0$ whose adjoint action preserves the subspace $\eta(V)$
of $\End_\R(S)$:
\be
\L(\eta)\eqdef \{a\in \Aut_\R(S_0)|\Ad(a)(\eta(V))=\eta(V)\}~~.
\ee
The {\em vector representation} of $\L(\eta)$ is the group morphism $\Ad_0: \L(\eta)\rightarrow \O(V,h)$ 
defined through:
\be
\Ad_0(a)\eqdef (\eta|_V)^{-1}\circ \Ad(a)|_{\eta(V)}\circ (\eta|_V)~~.
\ee 
A {\em real Lipschitz structure of type $\eta$} on $(M,g)$ is an
$\Ad_0$-reduction $(Q,\tau)$ of the principal bundle $P(M,g)$ of
pseudo-orthogonal frames of $(TM,g)$, i.e. a pair formed of a
principal $\L(\eta)$-bundle $Q$ over $M$ and an $\Ad_0$-equivariant
fiber bundle map $\tau:Q\rightarrow P(M,g)$.  A bundle $(S,\gamma)$ of
Clifford modules over $(M,g)$ is weakly-faithful iff each fiberwise
Clifford representation $\gamma_p:\Cl(T_pM,g_p)\rightarrow \End(S_p)$
(where $p\in M$) is weakly-faithful. Since $M$ is connected, all
fiberwise Clifford representations $\gamma_p$ are {\em unbasedly}
isomorphic \footnote{This means that they are isomorphic in a certain
  category which is defined in \cite{Lipschitz} and which has more
  morphisms than the usual category of representations. } with each
other and hence with some fiducial weakly faithful Clifford
representation $\eta:\Cl(V,h)\rightarrow \End_\R(S_0)$, where $S_0$ is
a vector space which models the fiber of $S$. The representation $\eta$
(considered up to unbased isomorphism of representations) is called
the {\em type} of $(S,\gamma)$. One has the following key result:

\

\noindent {\bf Theorem 1. \cite{Lipschitz}.}  There exists an
equivalence of categories between the groupoid of real Lipschitz
structures of type $\eta$ and the groupoid of real pinor bundles of
type $\eta$ defined over $(M,g)$. In particular, the underlying vector
bundle $S$ of every real pinor bundle $(S,\gamma)$ of type $\eta$ is
associated to the principal bundle $Q$ of a Lipschitz structure
$(Q,\tau)$ which has type $\eta$.

\

\noindent The theorem implies that $(M,g)$ admits a real pinor bundle
of type $\eta$ iff it admits a real Lipschitz structure of type
$\eta$, and that the classifications of these two kinds of objects up
to the corresponding notion of isomorphism agree.

One can show that any irreducible real Clifford representation is
weakly-faithful and that all such representations of $\Cl(V,h)$ belong
to the same {\em unbased} isomorphism class, which is determined by
the signature $(p,q)$ of $(V,h)$. A real pinor bundle $(S,\gamma)$ is
called {\em elementary} if its fibers are {\em irreducible} as real
Clifford representations, which amounts to the requirement that its
type $\eta$ is irreducible. The real Lipschitz groups of irreducible
Clifford representations are called {\em elementary}, as are the real
Lipschitz structures whose type is given by such representations. For
each quadratic vector space $(V,h)$, there exists an essentially
unique elementary real Lipschitz group $\L$, determined up to
isomorphism by the signature $(p,q)$ of $(V,h)$. Moreover, the nature
of this group depends only on $p-q \mod 8$. One can show that
$\L=\R_{>0}\times \cL$, where $\cL$ is a natural subgroup called the
{\em reduced Lipschitz group}, which can be constructed using the
so-called ``Lipschitz norm''. Elementary real Lipschitz groups were
classified in \cite{Lipschitz}. The result is summarized in Table
\ref{elementaryL}. A {\em reduced elementary Lipschitz structure} is
defined like a Lipschitz structure, but using the group $\cL$ (and the
restriction of $\Ad_0$ to $\cL$) instead of $\L$. The groupoid of
elementary real Lipschitz structures is equivalent with that of
reduced elementary real Lipschitz structures, so the latter is also
equivalent with the groupoid of elementary real pinor bundles.
When $pq\neq 0$, $\cL$ is neither compact nor connected. 

\begin{table}[H]
\label{elementaryL}
\centering
\begin{tabular}{|c|c|c|c|c|c|c|c|c|}
\hline $\begin{array}{c} p-q\\ {\rm mod}~8 \end{array}$ & $\cL$ & $\fG(p,q)$\\ 
\hline\hline                
$0,2$ & $\Pin(p,q)$ & $1$ \\ 
\hline 
$3,7$ & $\Spin^o(p,q)\eqdef \Spin(p,q)\cdot \Pin_2^{\alpha_{p,q}}$ & $\O(2,\R)$ \\ 
\hline $4,6$ & $\Pin^q(p,q)\eqdef \Pin(p,q)\cdot \Sp(1)$ & $\SO(3,\R)$ \\ 
\hline $1$ & $\Spin(p,q)$ & $1$ \\ 
\hline $5$ & $\Spin^q(p,q)\eqdef \Spin(p,q)\cdot \Sp(1)$ & $\SO(3,\R)$ \\ 
\hline
\end{tabular}
\vskip 0.2in
\caption{Reduced elementary Lipschitz groups in signature $(p,q)$. The
  sign factor $\alpha_{p,q}$ equals $-1$ when $p-q\equiv_8 3$ and $+1$
  when $p-q\equiv_8 7$ and we use the notation $\Pin_2^+\eqdef
  \Pin(2,0)$ and $\Pin_2^-\eqdef \Pin(0,2)$. The last column lists the
  characteristic group. The symbol ``$\cdot$'' denotes direct product
  of groups divided by a central $\Z_2$ subgroup.}
\end{table}
It is clear from this table that the conditions for existence of an
elementary Lipschitz structure are generally {\em weaker} 
(and sometimes considerably so) than those
for existence of a spin structure. Every elementary Lipschitz group
has a so-called {\em characteristic representation}, which is
naturally associated to it as explained in \cite{Lipschitz}. The image
of this representation is the so-called {\em characteristic group}
$\fG(V,h)$, whose isomorphism type is listed in the last column of
Table \ref{elementaryL}. Accordingly, an elementary Lipschitz
structure $Q$ induces a principal \emph{characteristic bundle} $E$
(with structure group $\fG(p,q)$), which is associated to $Q$ through
the characteristic representation of the corresponding Lipschitz
group; this bundle can be non-trivial only when $p-q\not \equiv_8
0,1,2$.  For $p-q\equiv_8 0,1,2$, a Lipschitz structure is either a
$\Spin$ or $\Pin$ structure and hence is of the classical type studied
for example in \cite{Karoubi}.  When $p-q\equiv_8 5$, it is a
$\Spin^q$ structure in general signature; the positive-definite case
($q=0$) of such was studied in \cite{Nagase}. The cases $p-q\equiv_8
4,6$ lead to $\Pin^q$ structures, which are a slight extension of
$\Spin^q$ structures to non-orientable manifolds. The cases
$p-q\equiv_8 3,7$ lead to what we call $\Spin^o$ structures (which
appear to be new).

The characteristic bundle of a $\Spin^{o}$-structure is a principal
$\O(2)$ bundle, which suggests that it may be relevant to situations where
spinors are \emph{charged} under a $\O(2)$ gauge group rather than
under a $\U(1)$ group. This fact may be relevant to understand the
worldvolume theories of non-orientable D-branes. Let: 
\beqa
\sigma:=\sigma_{p,q} &\eqdef& (-1)^{q+\left[\frac{d}{2}\right]}=
\twopartdef{(-1)^{\frac{p-q}{2}}}{d=\mathrm{even}}{(-1)^{\frac{p-q-1}{2}}}{d=\mathrm{odd}}=\nn\\
&=& \twopartdef{+1}{p-q\equiv_4 0,1}{-1}{p-q\equiv_4 2, 3}~~.
\eeqa
Let $\w_1^\pm(M)$ be the modified Stiefel-Whitney classes of $(M,g)$
introduced in \cite{Karoubi}; these classes depend on $g$ 
but we don't indicate this in the notation. The topological obstructions to
existence of elementary real Lipschitz structures (and hence of
elementary real pinor bundles) on $(M,g)$ are as follows
\cite{Lipschitz}:
\begin{itemize}
\item In the normal simple case ($p-q\equiv_{8} 0, 2$), $(M,g)$ admits
  an elementary real pinor bundle iff $(M,-\sigma g)$ admits a $\Pin$
  structure, which requires that the following condition is satisfied:
\begin{equation*}
\w_2^+(M)+\w_2^-(M) + \w^{\sigma}_1(M)^2+ \w^{-}_1(M) \w^{+}_{1}(M) =0\, .
\end{equation*}
\item In the complex case ($p-q\equiv_{8} 3, 7$), $(M,g)$ admits an
  elementary real pinor bundle iff it admits a
  $\Spin^{o}$-structure, which happens iff there exists a principal
  $\O(2,\R)$-bundle on $M$ such that the following two conditions are
  satisfied:
\be
\w_1(M) = \w_1(E) 
\ee
\beqa
\w_2^+(M)&+&\w_2^-(M)\!=\! \w_2(E)+ \w_1(E)(p \w_1^+(M)+q \w_1^-(M))\\
&+&\left[\delta_{\alpha,-1}+\frac{p(p+1)}{2}+\frac{q(q+1)}{2}\right]\w_1(E)^2~,
\eeqa
where $\alpha\eqdef \alpha_{p,q}$. 
\item In the quaternionic simple case ($p-q\equiv_{8} 4, 6$), $(M,g)$
  admits an elementary real pinor bundle iff $(M,-\sigma g)$ admits a
  $\Pin^{q}$-structure, which happens iff there exists a principal
  $\SO(3,\R)$-bundle $E$ on $M$ such that the following condition is
  satisfied:
\begin{equation*}
\w_2^+(M)+ \w_2^-(M)+\w^{\sigma}_1(M)^2+ \w^{-}_1(M) \w^{+}_{1}(M) = \w_2(E)~~.
\end{equation*}
\item In the normal non-simple case ($p-q\equiv_{8} 1$), $(M,g)$
  admits an elementary real pinor bundle iff it admits a $\Spin$
  structure, which requires that the following two conditions are
  satisfied:
\begin{equation*}
\label{eq:spinobs}
\w_1(M)=0\, , \qquad \w_2^+(M)+\w_2^-(M)= 0\, .
\end{equation*}	
\item In the quaternionic non-simple case ($p-q\equiv_{8} 5$), $(M,g)$
  admits an elementary real pinor bundle iff it admits a
  $\Spin^{q}$-structure, which happens iff there exists a principal
  $\SO(3,\R)$-bundle $E$ over $M$ such that the following conditions
  are satisfied:
\begin{equation*}
\w_1(M)=0\, , \qquad \w_2^+(M) +\w_2^-(M)= \w_2(E)\, .
\end{equation*}
\end{itemize}

\section{A\lowercase{pplications to} \lowercase{string} \lowercase{theory and} \lowercase{supergravity}}
\label{sec:Applications}

The results of reference \cite{Lipschitz} can be applied to study the
spinorial structures needed to formulate various supergravity
theories.  In this section, we sketch a simple application to
M-theory, obtaining a no-go result regarding the global interpretation
of its spinor fields.

Consider M-theory on an eleven-dimensional Lorentzian manifold of
``mostly plus'' signature $(p,q)=(10,1)$. The low energy limit is
given by eleven-dimensional supergravity, whose supersymmetry
generator is a $32$-component real spinor $\epsilon$. The gravitino
Killing spinor equation contains terms with an odd number of gamma
matrices acting on $\epsilon$, implying that the whole Clifford
algebra $\Cl(T_xM,g_x)$ at a point $x\in M$ must act on the value of
$\epsilon$ at $x$. If one assumes that $\epsilon$ is a global section
of a vector bundle $S$ endowed with {\em inner} Clifford multiplication, it
follows that each fiber $S_p$ must carry a real irreducible
representation of $\Cl(T_pM,g_p)$ and hence that $S$ is an elementary
real pinor bundle.  Since $p-q=9\equiv_8 1$, we are in the normal
simple case. Hence $(M,g)$ admits an elementary real pinor bundle $S$
if and only if it is oriented and spin. Since $\w_2^-(M) = 0$, the
corresponding topological obstruction can be written as follows:
\be
\w^{+}_1(M)= \w^{-}_{1}(M)\, , \qquad \w^{+}_2(M)=0\, .
\ee
We conclude that, in signature $(10,1)$, the supersymmetry parameter
can be interpreted as a global section of an elementary real pinor
bundle iff the space-time is orientable and spin.

Of course, M-theory can in fact be defined on Lorentzian
eleven-manifolds admitting a Pin structure
\cite{WittenFluxQuantization,WittenParityAnomaly}, but that
construction involves a bundle with external Clifford multiplication,
which leads to a modified Dirac operator as in \cite{ModifiedDirac}.

\section{F\lowercase{uture directions}}
\label{sec:Future}

The results of \cite{Lipschitz} open up various directions for further
research.  Here are some questions which may be worth pursuing:

\begin{itemize}
\item Reference \cite{Lipschitz} classifies bundles of irreducible
  modules over $\Cl(M,g)$. It would be interesting to classify bundles
  of faithful real Clifford modules over $\Cl(M,g)$ and irreducible or
  faithful real Clifford modules over the even sub-bundle
  $\Cl^\ev(M,g)$, since such bundles may also be relevant in string
  theory and supergravity.
\item It would be interesting to study the index theorem for general
  bundles of real Clifford modules, without assuming that $(M,g)$ is
  spin.
\item One could consider extending Wang's classification
  \cite{Wangparallel} beyond the case of spin manifolds,
  characterizing manifolds admitting sections of an elementary real
  pinor bundle which are parallel with respect to a connection lifting
  the Levi-Civita connection on $(M,g)$ and a fixed connection on the
  characteristic bundle.
\item Killing and generalized Killing spinors were studied in the
  literature \cite{Bar,2013arXiv1303.6179M,MoroianuSpinc}, usually on
  manifolds carrying a fixed $\Spin$ or $\Spin^{c}$ structure. Using
  our results, this could be extended to the most general
  pseudo-Riemannian manifolds admitting elementary real pinor bundles.
\item One could apply our results to the spin geometry of branes in
  string and M-theory. As shown in reference \cite{Freed:1999vc}, the
  worldvolume of orientable D-branes in the absence of $H$-flux admits
  a $\Spin^{c}$-structure. In the unorientable case, this may become a
  Lipschitz structure.
 \item Our results may be useful to globally characterize the local spinor bundles appearing in exceptional generalized geometry \cite{Pacheco:2008ps}, obtaining the topological obstructions to their existence.
\end{itemize} 

\begin{acknowledgments}  
The work of C.I.L. is supported by grant IBS-R003-S1. The work of
C.S.S. is supported by the ERC Starting Grant 259133 – Observable
String.
\end{acknowledgments}

\appendix


\end{document}